\documentclass[aps,prb,preprint,superscriptaddress,amsmath,amssymb,onecolumn]{revtex4-2}
\usepackage{graphicx}
\usepackage{dcolumn}
\usepackage{bm}
\usepackage{multirow}
\usepackage{lineno}
\usepackage[colorlinks,linkcolor=blue,anchorcolor=blue,citecolor=blue]{hyperref}
\usepackage{color}

\begin{document}
\title{Robust zero-energy states in two-dimensional Su-Schrieffer-Heeger topological insulators}

\author{Zhang-Zhao Yang}
\author{An-Yang Guan}
\author{Wen-Jie Yang}
\affiliation{Key Laboratory of Modern Acoustics, MOE, Institute of Acoustics, Department of Physics, Collaborative Innovation Center of Advanced Microstructures, Nanjing University, Nanjing 210093, People’s Republic of China}

\author{Xin-Ye Zou}
\email{xyzou@nju.edu.cn}
\author{Jian-Chun Cheng}
\affiliation{Key Laboratory of Modern Acoustics, MOE, Institute of Acoustics, Department of Physics, Collaborative Innovation Center of Advanced Microstructures, Nanjing University, Nanjing 210093, People’s Republic of China}
\affiliation{State Key Laboratory of Acoustics, Chinese Academy of Sciences, Beijing 100190, People’s Republic of China}

\begin{abstract}
The Su-Schrieffer-Heeger (SSH) model on a two-dimensional square lattice has been considered as a significant platform for studying topological multipole insulators. However, due to the highly-degenerate bulk energy bands protected by $ C_{4v} $ and chiral symmetry, the discussion of the zero-energy topological corner states and the corresponding physical realization have been rarely presented. In this work, by tuning the hopping terms to break $ C_{4v} $ symmetry down to $ C_{2v} $ symmetry but with the topological phase invariant, we show that the degeneracies can be removed and a complete band gap can be opened, which provides robust protection for the spectrally isolated zero-energy corner states. Meanwhile, we propose a rigorous acoustic crystalline insulator and therefore these states can be observed directly. Our work reveals the topological properties of the robust zero-energy states, and provides a new way to explore novel topological phenomena.
\end{abstract}

\maketitle

\section{INTRODUCTION}
The concept of higher-order topological (HOT) phases arising from the modern theory of charge polarization in solids have greatly extended the class of the topological insulators (TIs) \cite{HOT6346,HOT245115,HOT106802,HOT246401,HOT246402,HOT0346,HOT245151,HOT033074}. Distinct from the bulk-boundary correspondence of the traditional TIs \cite{TI1,TI2,TI3,TI4,TI5}: the Chern insulators that require breaking time-reversal symmetry \cite{CI1,CI2,CI3,Chec1,Chec2,Chec3,Chec4,Chec5,Chec6,Chec7,Chec8,Chec9,Chec10,Chec11} and the $ Z_{2} $ insulators that are time-reversal invariant \cite{QSH1,QSH2,QSH3,Z2c1,Z2c2,Z2c3,Z2c4,Z2c5,Z2c6,Z2c7,Z2c8,Z2c9,Z2c10,Z2c11,Z2c12,Z2c13,Z2c14}, these HOT insulators characterized by quantized nontrivial bulk polarization are protected by intrinsic crystalline symmetries and can induce the filling anomalies of the fractional charges at the boundaries or corners of the crystal structures \cite{HOT106802,HOT245151,Slager1,Slager2,Filling}. Recently, the HOT insulators have been theoretically and experimentally demonstrated in circuits \cite{CC1,CC2}, microwaves \cite{MW1}, optics \cite{HOTOP1,HOTOP2,2DSSH,HOTOP3,HOTOP4} and acoustics \cite{HOTAC1,YZZ,HOTAC2,HOTAC3,HOTAC4,HOTAC5,HOTAC6}. 

Generally, the observation of the obstructed topological states induced by the charge fractionalization always requires spectral isolation, for that the robust corner states are at the mid-gap if particle-hole symmetry or chiral symmetry is preserved \cite{HOTOP2}. Otherwise, these topological states may degenerate with the trivial bulk modes, and then act as symmetry-protected bound states in the continuum \cite{20BIC1,20BIC2}. Especially, for the lattices with $ C_{4v} $ symmetry as well as chiral symmetry, the double degenerate bands at the high symmetry points $ M $ and $ \Gamma $ always result in the gap being closed \cite{Liufeng1,Liufeng2}, which naturally hinders the identification of the zero-energy topological corner states. Previous works on $ C_{4v} $-symmetric classical systems have presented the domain-wall HOT states located in the lower gap that is characterized by a dipole moment \cite{HOTOP4}, but the discussion on the zero-energy corner states in such the systems and the corresponding physical realization still need to be addressed. 

In this paper, we demonstrate the existence of the isolated zero-energy topological corner states in a two-dimensional (2D) Su-Schrieffer-Heeger (SSH) model, and present the corresponding realization in acoustic system. By judiciously breaking $ C_{4v} $ symmetry with the preservation of all the other certain symmetries, i.e., $ C_{2v} $ symmetry and chiral symmetry, which ensure the topological phase invariant during the process, the degeneracy of the modes at the high symmetry points can be removed, and the zero-energy topological corner states can exist at the mid-gap. Meanwhile, we show that these zero-energy states are so robust against perturbations even if all symmetries are reduced. According to these results, we then numerically propose an acoustic HOT insulator which rigorously corresponds to the presented 2D SSH model, and all the theoretical predictions can be directly observed. 

This paper is organized as follows. The 2D SSH theoretical model along with its topological states is introduced in Sec. \hyperref[sec2]{II}. In Sec. \hyperref[sec3]{III}, the discussion of the robustness of the zero-energy topological corner states is presented. In section \hyperref[sec4]{IV}, the corresponding acoustic HOT insulator is proposed. Finally, a summary is given in Sec. \hyperref[sec5]{V}. A few appendixes are provided as a supplement to the discussion in the main text.

\section{TWO-DIMENSIONAL SSH MODEL WITH ZERO-ENERGY TOPOLOGICAL STATES \label{sec2}}
\begin{figure}
    \centering
    \includegraphics[width=0.95\textwidth]{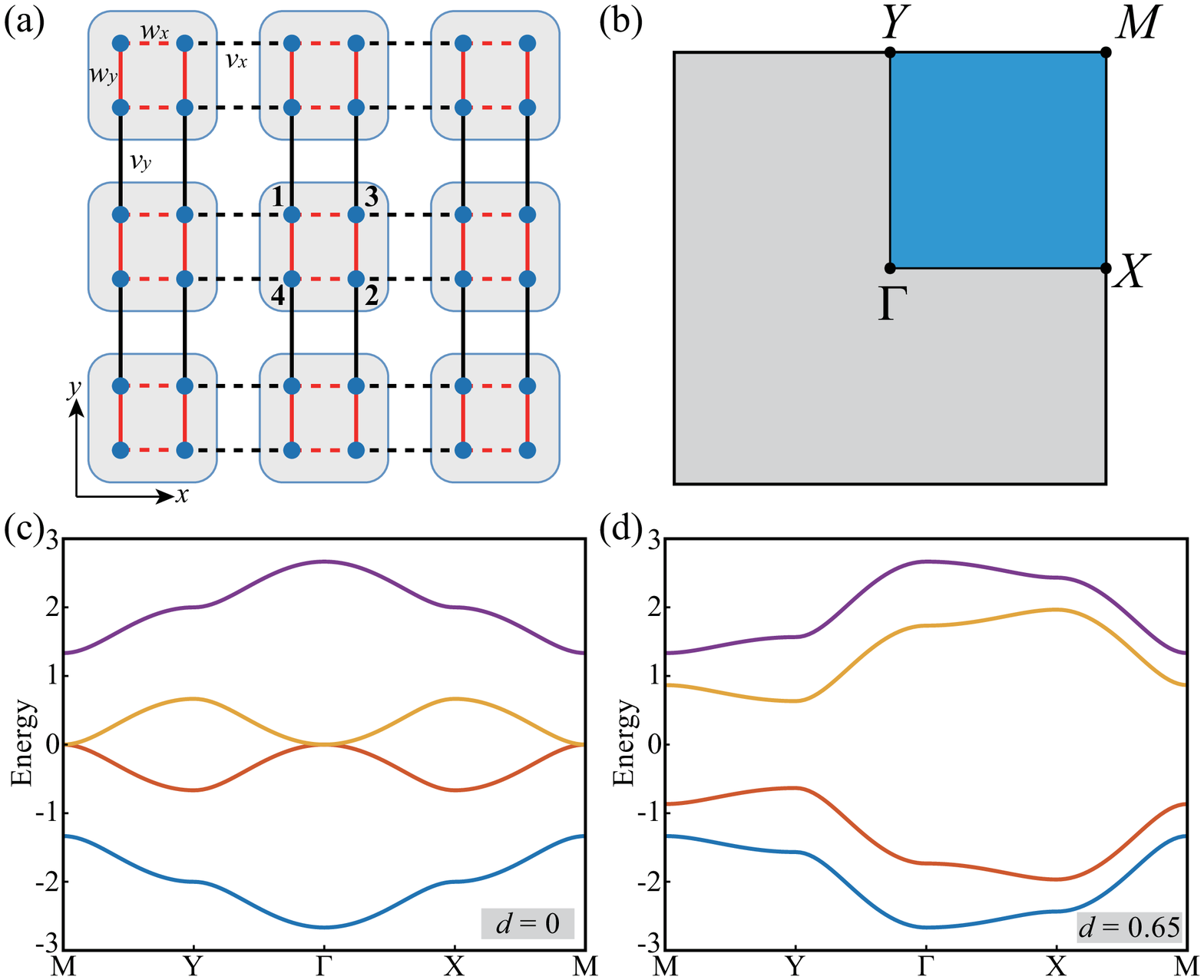}
    \caption{(a) Schematic of the 2D SSH model. (b) First BZ of the square lattice. Energy band structures of the bulk lattice when (c) $ d = 0 $ and (d) $ d = 0.65 $.}
    \label{fig1}
\end{figure}

In this section, we present the topological properties of the 2D SSH model. As shown in Fig. \hyperref[fig1]{1(a)}, there are four atomic sites (labeled with 1-4, respectively) within the square lattice. Here, we neglect the impact of the onsite energy and only consider the nearest-neighbor hopping. The corresponding first Brillouin zone (BZ) is depicted in Fig. \hyperref[fig1]{1(b)}, and the Hamiltonian of this model can be given as
\begin{equation}
\begin{aligned}
h({\rm \textbf{k}}) = & -(w_{x}+v_{x}{\rm cos} k_{x})\tau_{x}\sigma_{0} + v_{x}{\rm sin} k_{x}\tau_{y}\sigma_{z}\\
& -(w_{y}+v_{y}{\rm cos}k_{y})\tau_{x}\sigma_{x} - v_{y}{\rm sin}k_{y}\tau_{x}\sigma_{y},
\end{aligned}
\label{eq1}
\end{equation}
where $ \textbf{k}=(k_{x},k_{y}) $ and $\{w_i,v_i,k_i\}(i=x,y)$ are the intra- and inter-lattice hopping terms, and the basis reciprocal vector in the $ i $-direction, respectively. $ \tau $ and $ \sigma $ are the Pauli matrix, while $ \sigma_{0} $ is the identity matrix. 

According to the theory of topological multipole insulators, the charges are predicted to accumulate at the boundaries of the lattice (we label the left (right) boundary as $ x $-edge, and the up (bottom) boundary as $ y $-edge, respectively, due to the translation invariance), as the total effect of the occupied bands acts as dipole moments in the corresponding direction. In 2D crystalline systems, the topological index can be characterized by the 2D Zak phase which is related to the charge polarization $\mathcal{P}=(P_{x},P_{y})$, where \cite{HOT6346,HOT245115}
\begin{equation}
    P_{j}=-\frac{1}{(2\pi)^{2}}\int_{1BZ}{\rm Tr}[\mathcal{A}_{j,\textbf{k}}]d^{2}\textbf{k},
    \label{eq2}
\end{equation}
where $ [A_{j,\textbf{k}}]^{mn}=-i\langle u_{\textbf{k}}^{m}\lvert \partial_{k_{j}}\lvert u_{\textbf{k}}^{n} \rangle$ $ (j=x,y) $ is the non-Abelian Berry connection (where $ \lvert u_{\textbf{k}}^{m}\rangle $ is the Bloch wave function and $ m $, $ n $ run over the occupied bands). Note that the two components $ P_{x} $ and $ P_{y} $ are independent with each other. On the other hand, when the lattice is $ C_{4v} $-symmetric, which also implies chiral symmetry, there are always degenerate bands existing at the zero energy of the high symmetry points $ M $ and $ \Gamma $ of the BZ. To lift the degenerates, we then introduce a set of basis parameters $ \{w,v,d\} $ and redefine the hopping terms as $ w_{x}=w(1-d) $, $ v_{x}=v(1-d) $ and $ w_{y}=w(1+d) $, $ v_{y}=v(1+d) $, respectively. Therefore, it is obvious to see that the charge polarization $\mathcal{P}$ is yielded between $(0,0)$ and $(1/2,1/2)$, and is purely determined by the hopping ratio $ \beta = v/w $; meanwhile, the nonzero $ d\in (-1,1) $ results in $ C_{4v} $ symmetry broken down to $ C_{2v} $ symmetry. As a result, if only one occupied band is considered, $ \beta > 1 $ and $ C_{2v} $ symmetry yield the polarization $ \mathcal{P}=(1/2,1/2) $, which represents the quantized dipole moments bounded at four boundaries; $ d>1/\beta $ indicates that the degenerates are totally removed and a complete band gap is opened at zero energy. Figure \hyperref[fig1]{1(d)} shows the bulk energy band structure when $ \{v,\beta,d\}=\{1,3,0.65\} $. As a comparison, the energy band structure when $ d = 0 $ is presented in Fig. \hyperref[fig1]{1(c)}. Further, we show the existence of the topological edge states and zero-energy corner states in this model. 

\begin{figure}
    \centering
    \includegraphics[width=0.95\textwidth]{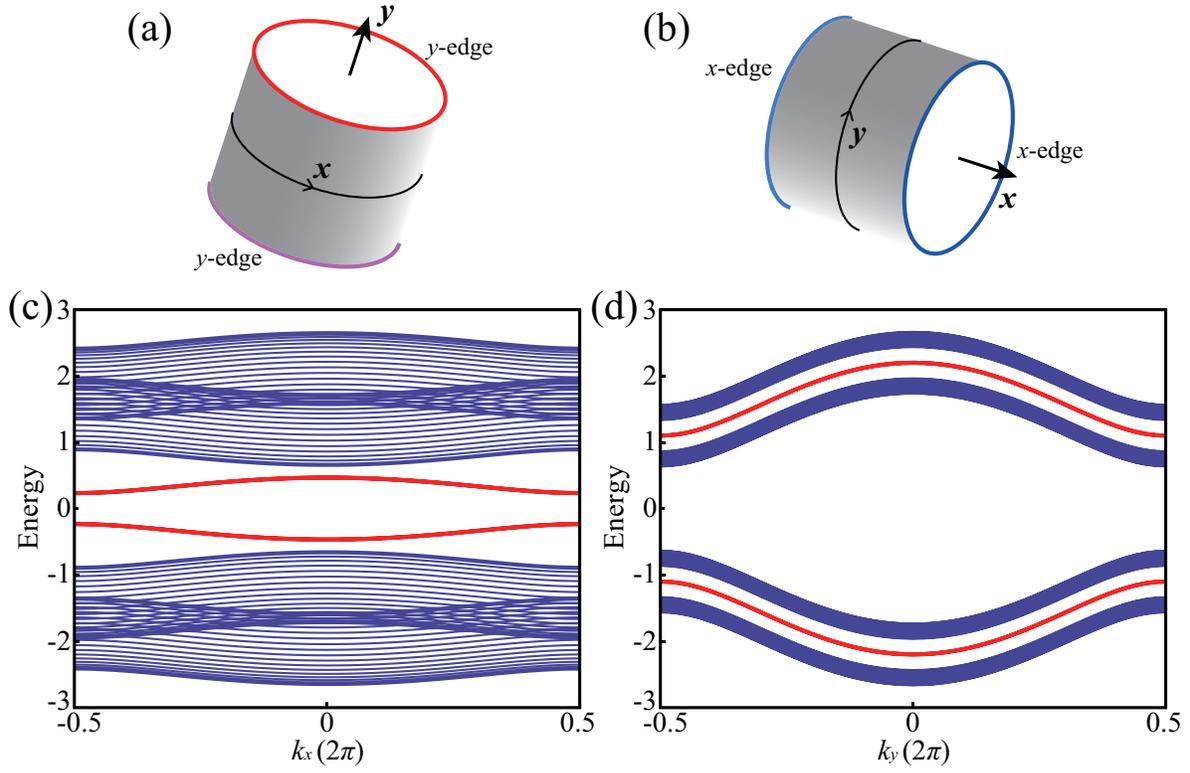}
    \caption{Schematics of tubular structures with (a) closed $ x $-edges and open $ y $-edges, and (b) open $ x $-edges and  closed $ y $-edges. (c) Energy band structure of the ribbon-shaped superlattice in (a). (d) Energy band structure of the ribbon-shaped supperlattice in (b).}
    \label{fig2}
\end{figure}

As discussed above, the topological edge-localized states are predicted to emerge along with the dipole moments. To explain this, we open the boundaries in the $y (x)$ direction while remaining the boundaries in the $x (y)$ direction periodic, and the corresponding quasi-one-dimensional tubular structures are illustrated in Figs. \hyperref[fig2]{2(a)} and \hyperref[fig2]{2(b)}, respectively. Figures \hyperref[fig2]{2(c)} and \hyperref[fig2]{2(d)} show the energy band structures of the ribbon-shaped superlattices in Figs. \hyperref[fig2]{2(a)} and \hyperref[fig2]{2(b)}, respectively. It can be seen that in both sides, the bands with edge-localized states are spectrally isolated from the bulk bands. It is worth noting that the dipole moments in this model are actually hosted by the first and fourth bands of the lattice, whereas the total polarization vanishes in the middle gap, for that there are two occupied bands below this gap. Therefore, the fact that the edge states at $ y $-edges are located in the middle gap is due to $ C_{4} $ symmetry broken that leads to the large offset of the bulk bands. Note that the same results can also be characterized by the Wannier bands as illustrated in Appendix. \hyperref[apdxA]{A}.

\begin{figure}
    \centering
    \includegraphics[width=0.95\textwidth]{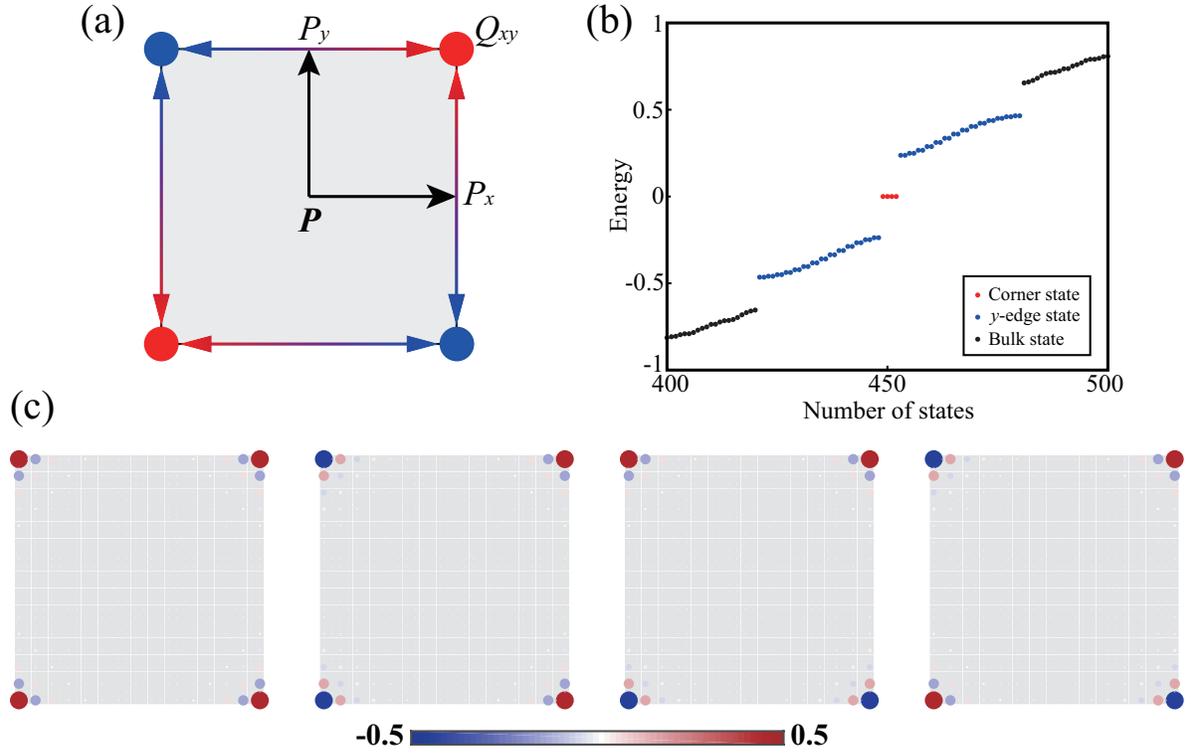}
    \caption{(a) Illustration of the polarizations in the 2D SSH model. (b) Energy spectrum of the 15$\times$15 finite structure. (c) Spatial energy distributions of the four degenerate zero-energy corner states.}
    \label{fig3}
\end{figure}

Although the total polarization in the middle gap vanishes, the independent components of the dipole moments still allow us to define a quadrupole index as \cite{Liufeng3}
\begin{equation}
    \mathcal{Q}_{xy}=\sum_{n=1}^{occ}P_{x}^{n}P_{y}^{n}.
    \label{eq3}
\end{equation}
For that there are two occupied bands for the middle gap, $ \mathcal{Q}_{xy}=1/2 $ indicates a double projection of the convergent dipole moments can exist at the corners of the system [illustrated in Fig. \hyperref[fig3]{3(a)}], which is also associated with the corner-induced filling anomaly. Here, we consider a finite structure with full open boundaries spanning $15\times 15$ lattices, and the corresponding calculated energy spectrum is presented in Fig. \hyperref[fig3]{3(b)}. It can be seen that in addition to the $ y $-edge-localized states, there are four degenerate zero-energy corner states protected by chiral symmetry emerging in the middle gap. The corresponding spatial energy distributions of the four degenerate corner states are presented in Fig. \hyperref[fig3]{3(c)}. Meanwhile, due to the fact that these corner states are isolated from the trivial bulk modes, all these states can carry $ 1/2 $ fractional charges and are exponentially confined to the corners.

\section{ROBUSTNESS OF THE ZERO-ENERGY CORNER STATES \label{sec3}}
\begin{figure}
    \centering
    \includegraphics[width=0.95\textwidth]{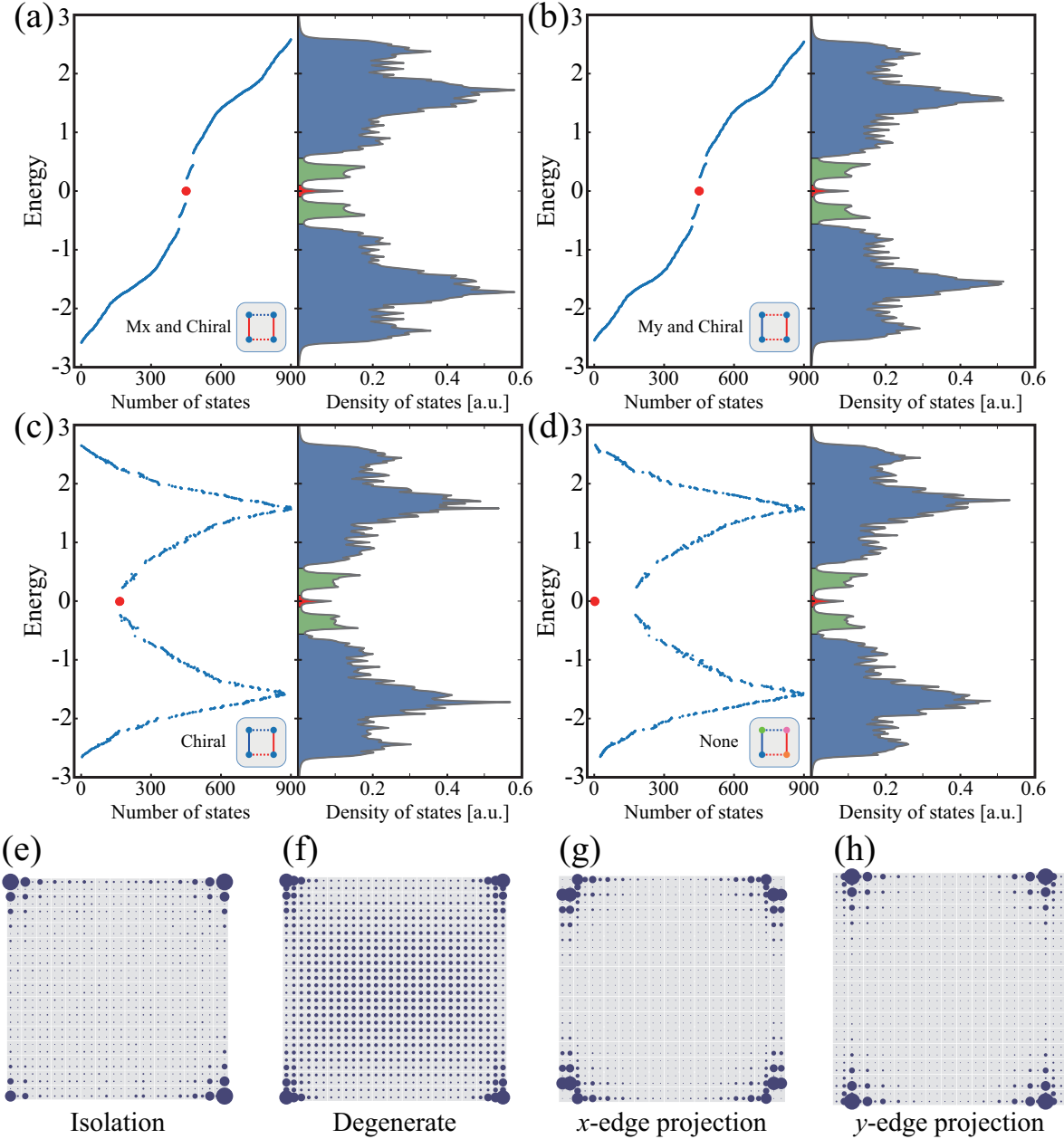}
    \caption{Energy spectra along with the corresponding denity functions of the perturbed structures with (a) $ M_x $ symmetry and chiral symmetry, (b) $ M_y $ symmetry and chiral symmetry, (c) only chiral symmetry and (d) none symmetry. Insets: Schematics of a lattice with distorted hopping terms or on-site terms. Spatial energy distributions of (e) isolated zero-energy corner states, (f) degenerate zero-energy corner states, (g) corner states projected by $ x $-edge dipole moments and (d) corner states projected by $ y $-edge dipole moments.}
    \label{fig4}
\end{figure}

In this section, we present the high robustness of these isolated states. As discussed above, apart from time-reversal symmetry, the 2D SSH model also possesses $ C_{2v} $ symmetry, which includes mirror symmetries $ M_{x} $ and $ M_{y} $, and chiral symmetry $ \Pi $. Here, we add a small perturbation $ h_{per}^{\Lambda} $ consisting of random hopping terms to the bulk lattices of the finite structure [Fig. \hyperref[fig3]{3(b)}] as
\begin{equation}
    h^{\Lambda}(\textbf{k}) = h(\textbf{k}) + h_{per}^{\Lambda},
    \label{eq4}
\end{equation}
where $ \Lambda =\{M_{x},M_{y},\Pi \} $ indicates that the perturbation respects the corresponding symmetry as $ [h_{per}^{\Lambda},\Lambda]=0 $. As a result, these well-tuned symmetries are broken down to only specific indicated symmetries \cite{HOTOP2,20BIC1}. In Figs. \hyperref[fig4]{4(a)-4(d)} we show the energy spectra along with the density functions of the perturbed structures, and it is obvious to see that the zero-energy states are so robust against perturbations which breaks mirror symmetries. In particular, even if all symmetries are broken (including chiral symmetry being slightly reduced), the topological corner states are still almost pinned to the zero energy and not degenerate with the bulk modes [Fig. \hyperref[fig4]{4(d)}].

We argue that the robust zero-energy states in the present 2D SSH model are actually protected by both the band gap and chiral symmetry \cite{20BIC1}. Whereas, once the introduced perturbations are strong enough to impact the dipole moments, these corner states may be unpinned and shifted to degenerate with the trivial bulk modes. In particular, the large distortion at the corners may result in the double-projections being removed, and the projection states which originate from the single dipole moment can be separated from the edge states and isolated to the corners. To explain this, the field distributions of the isolated and degenerate zero-energy states are shown in Figs. \hyperref[fig4]{4(e)} and \hyperref[fig4]{4(f)}, while the single-dipole-projected states are presented in Figs. \hyperref[fig4]{4(g)} and \hyperref[fig4]{4(h)}, respectively.

\section{ACOUSTIC REALIZATION OF THE 2D SSH MODEL \label{sec4}}
The proceeding discussion has theoretically demonstrated the existence of robust zero-energy topological corner states isolated in the mid-gap. In this section, we propose a rigorous physical model based on acoustic resonance system to observe these topological states directly. Figure \hyperref[fig5]{5(a)} provides the schematic of the crystalline structure that spans $ 10\times 10 $ lattice according to the model presented in Fig. \hyperref[fig3]{3(b)}. As illustrated, there are four identical cubic acoustic cavities connected by waveguide tubes in one lattice. The side length of the cavity is $ s $ = 6 cm, and the radius of the tube is $ r $ = 0.5 cm. For the bulk lattices, the lengths of the intra- and inter-lattices tubes in the two directions are $ l_{x}^{w} $ = 15.8 cm, $ l_{y}^{w} $ = 2.7 cm, $ l_{x}^{v} $ = 4.7 cm and $ l_{x}^{v} $ = 0.3 cm, respectively; for the lattices at the boundaries, all the outermost tubes need a correction length of $ 0.85r $ \cite{add2}. The mass density of the air and the corresponding sound speed are $ \rho_{0} $ = 1.29 ${\rm kg/m^{3}}$  and $ c_{0} $ = 343 m/s, respectively. Therefore, the bulk lattice of the acoustic structure then can be rigorously described by the theoretical model given in Eq. \hyperref[eq1]{(1)}. The corresponding hopping terms of the Hamiltonian can be calculated as $w_{x}=2.56\times 10^{5}$, $w_y=1.2\times 10^{6}$ and $ \beta = 3 $, respectively, and the on-site terms can be obtained as $\omega_{0}^{2} = w_{x}+w_{y}+v_{x}+v_{y} $ (see detailed derivation in Appendix \hyperref[apdxB]{B}). Accordingly, the numerical and theoretical results of the bulk energy bands are shown in Fig. \hyperref[fig5]{5(b)}, and the corresponding four eigenstates at the high symmetry points $X$ and $Y$ of the acoustic structure are presented in Figs. \hyperref[fig5]{5(c)} and \hyperref[fig5]{5(d)}, respectively. 

Further, we apply the acoustic absolute soft boundary condition, which guarantees the Hermiticity of the system and yields the acoustic wave at the boundaries to be zero, to the outermost tubes of the finite structure as a compensation to the on-site potential of the lattices at the boundaries \cite{add1,add2}, and the calculated energy band structures of the two ribbon-shaped superlattices consisting of 15 bulk lattices in the $ x $-direction and the $ y $-direction are presented in Figs. \hyperref[fig6]{6(a)} and \hyperref[fig6]{6(b)}, respectively, which corresponds to the theoretical results shown in Figs. \hyperref[fig2]{2(c)} and \hyperref[fig2]{2(d)}. For the finite structure with open boundaries [Fig. \hyperref[fig5]{5(a)}], the calculated eigenfrequency spectrum is then presented in Fig. \hyperref[fig6]{6(c)}. To verify the existence of acoustic higher-order topological states, the sound pressure distributions of the bulk state, edge state and corner state are provided in Fig \hyperref[fig6]{6(d)}. Such the results are perfectly corresponding to the theoretical predictions. In particular, the frequency of the acoustic corner states is 393 Hz, which is consistent with the theoretical prediction as $ f_{0}=\omega_{0}/2\pi $ = 385 Hz.

\section{SUMMARY\label{sec5}}
In summary, we have demonstrated the existence of the robust zero-energy topological corner states in the 2D SSH model, and directly observed these states based on a judiciously designed acoustic crystalline insulator. By breaking $ C_{4v} $ symmetry down to $ C_{2v} $ symmetry without impacting the topological properties of the bulk, the degeneracy of the bulk bands can be continuously removed, and result in a complete band gap opened at zero energy. Meanwhile, we show that with the protection of the band gap, these zero-energy corner states can be exponentially confined to the corners and are robustly against random perturbations introduced within the system. We also propose a rigorous acoustic topological crystalline insulator to verify these states directly. Our findings are expected to not only be helpful for the understanding of HOT states in topological multipole insulators, but also provide a platform for the design and potential applications of topological materials. 

\begin{acknowledgments}
Z.-Z. Y thanks Si-Ping Song for helpful discussions. This work was supported by the National Key R\&D Program of China (Grant No. 2017YFA0303700), National Natural Science Foundation of China (Grant Nos. 11634006, 11934009, and 12074184), the Natural Science Foundation of Jiangsu Province (Grant No. BK20191245), State Key Laboratory of Acoustics, Chinese Academy of Sciences.
\end{acknowledgments}

\appendix
\section{WANNIER BANDS OF THE 2D SSH MODEL\label{apdxA}}
We briefly introduce the Wilson-loop method and the corresponding results of the 2D SSH model. Here, we consider two occupied bands, and the discretized Berry connection matrix in the $ x $ and $ y $ directions can be defined as $ [F_{i,\textbf{k}}]^{mn} = \langle u_{\textbf{k}+\Delta k_{i}}^{m}\lvert u_{\textbf{k}}^{n}\rangle $ $ (i=x,y) $, where $ \Delta k_{i} $ is a small positive quantity of $ k $ in the $ i $-direction. The corresponding Wilson-loop operator then can be defined as $ \mathcal{W}_{i,\textbf{k}}=\Pi_{t_{i} = 0}^{N-1} F_{i,\textbf{k}+t\Delta k_{i}} $, where $ t_{i} = 2\pi/\Delta k_{i} $ is the number of the points that the loop in the $ i $ direction is discretized into. With fully periodic boundary conditions, $ \mathcal{W}_{x,\textbf{k}} $ can be diagonalized as 
\begin{equation}
    \mathcal{W}_{x,\textbf{k}}=\sum_{j}\lvert \nu_{x,\textbf{k}}^{j}\rangle e^{i2\pi\nu_{x}^{j}(k_{y})}\langle \nu_{i,\textbf{k}}^{j}\lvert,
    \label{eqA1}
\end{equation}
where $ \lvert \nu_{x,\textbf{k}}^{j}\rangle $ is the eigenstates and $ \nu_{x}^{j}(k_{y}) $ gives the Wannier centers of the two occupied bands in the $ k_{x} $-direction \cite{HOT245115}. By traversing the loop over the entire BZ along the $ k_{y} $-direction, we then can obtain two Wannier bands that are related to the polarization of the Bloch bands. The Wilson-loop operators in the $ k_{y} $-direction $ \mathcal{W}_{y,\textbf{k}} $ can be calculated similarly, yielding the Wannier bands as $ \nu_{y}^{j}(k_{x}) $. For a single occupied band, the quantized Wannier bands of the 2D SSH model are presented in Figs. \hyperref[fig7]{7(a)} and \hyperref[fig7]{7(b)}, respectively, which represents the existence of dipole moments on both $ x $-edges and $ y $-edges. The polarizations of the two occupied bands case are depicted in Figs. \hyperref[fig7]{7(c)} and \hyperref[fig7]{7(d)}, respectively. It is obvious to see that the two Wannier bands are quantized to be zero for the loops in the $ k_{x} $-direction, and the Wannier bands in the $ k_{y} $-direction are 1/2. This result implies that the charges can still accumulate on the $ y $-edges, but disappear on the $ x $-edges, which exactly corresponds to the results shown in Figs. \hyperref[fig2]{2(c)} and \hyperref[fig2]{2(d)}, respectively.

In addition, even if the total bulk polarization of the two occupied bands vanishes, the charges can still accumulate on the corners \cite{HOT245115}. This process can be discribed by the nested Wilson loops. Due to the fact that the Wannier bands are gapped in this model, the two bands can be labeled with $ ``\pm" $, respectively. The Wannier band subspaces then can be defined as
\begin{equation}
    \lvert w_{i,\textbf{k}}^{\pm}\rangle = \sum_{n=1}^{occ}\lvert u_{\textbf{k}}^{n}\rangle [\nu_{i,\textbf{k}}^{\pm}]^{n}.
    \label{eqA2}
\end{equation}
The nested discretized Berry connection then can be defined as $ F_{i,\textbf{k}}^{\pm} = \langle w_{\textbf{k}+\Delta k_{i}}^{\pm}\lvert w_{\textbf{k}}^{\pm}\rangle $, and the corresponding nested Wilson loops can be obtained as $\tilde{\mathcal{W}}_{i,\textbf{k}}^{\pm}=\Pi_{t_{i} = 0}^{N-1} F_{i,\textbf{k}+t\Delta k_{i}}^{\pm} $. Finally, the associated polarizations can be obtained as
\begin{equation}
    p_{y}^{\nu_{x}^{\pm}}=-\frac{i}{2\pi}{\rm Log}[\tilde{\mathcal{W}}_{y,k_{x}}^{\pm}],
    \label{eqA3}
\end{equation}
and the same for $ p_{x}^{\nu_{y}^{\pm}} $.

\section{DETAILED DERIVATION OF THE ACOUSTIC CRYSTALLINE MODEL\label{apdxB}}
\begin{figure}
    \centering
    \includegraphics[width=0.95\textwidth]{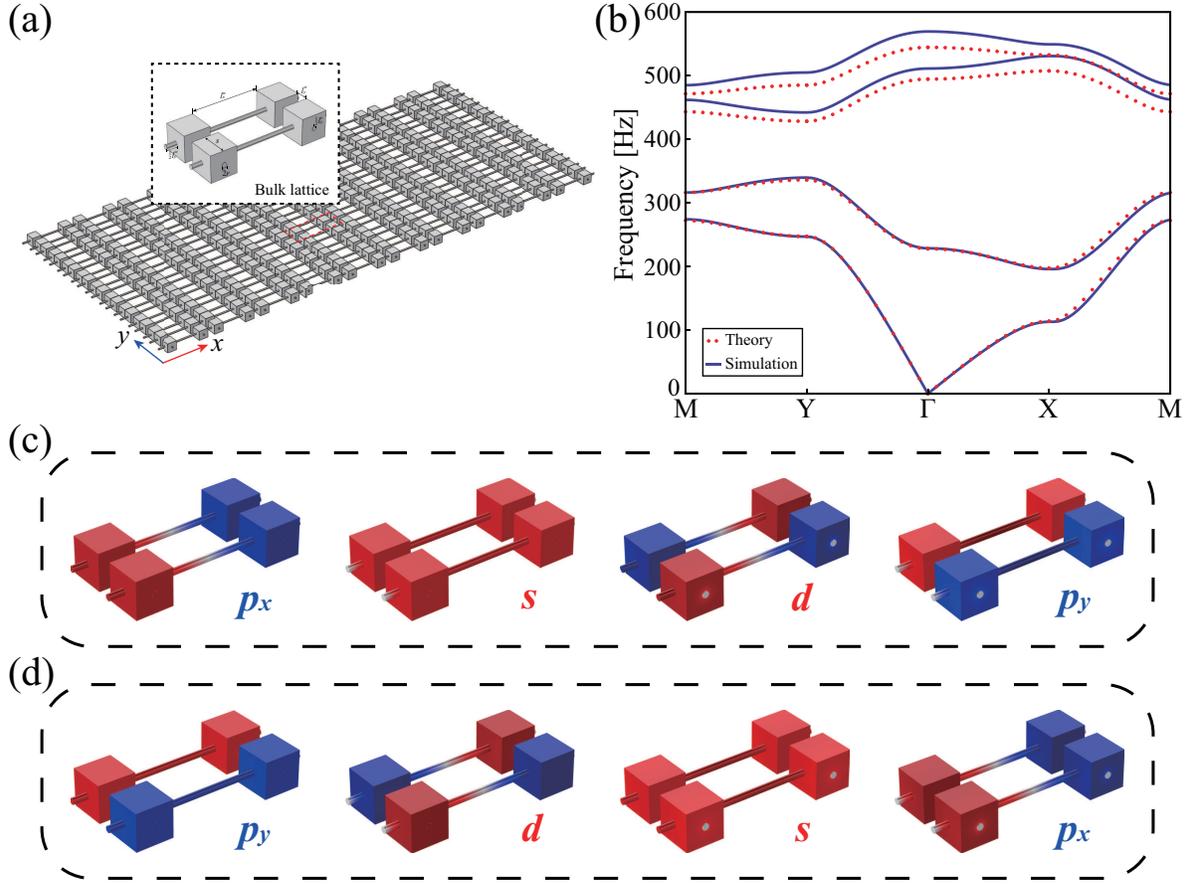}
    \caption{(a) Schematic of the finite acoustic crystalline insulator spanning $10 \times 10$ lattices. Inset: Bulk lattice of the acoustic structure. (b) Energy band structures of the acoustic bulk lattice calculated by simulation (blue lines) and theory (red dots), respectively. Four eigenstates from the bottom band to the top band of the high symmetry points (c) $ X $ and (d) $ Y $, respectively.}
    \label{fig5}
\end{figure}

\begin{figure}
    \centering
    \includegraphics[width=0.95\textwidth]{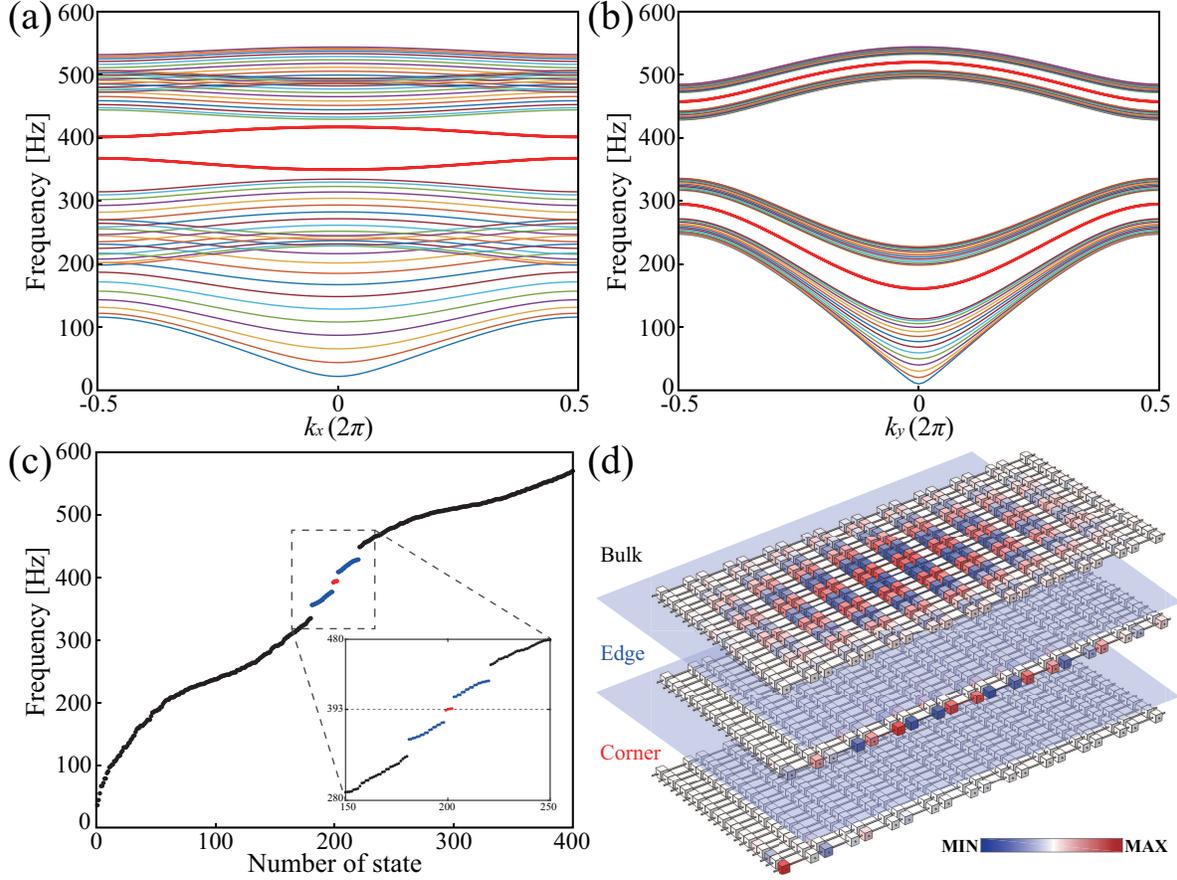}
    \caption{Energy band structures of the acoustic ribbon-shaped superlattices with (a) periodic $ x $-edges and open $ y $-edges, and (b) open $ x $-edges and  periodic $ y $-edges. Both the two superlattices host topological edge states (isolated red lines). (c) Numerically calculated eigenfrequency spetrum of the acoustic structure in Fig. 5(a). (d) Sound pressure distributions of the bulk state, $ y $-edge state and zero-energy corner state.}
    \label{fig6}
\end{figure}

\begin{figure}
    \centering
    \includegraphics[width=0.95\textwidth]{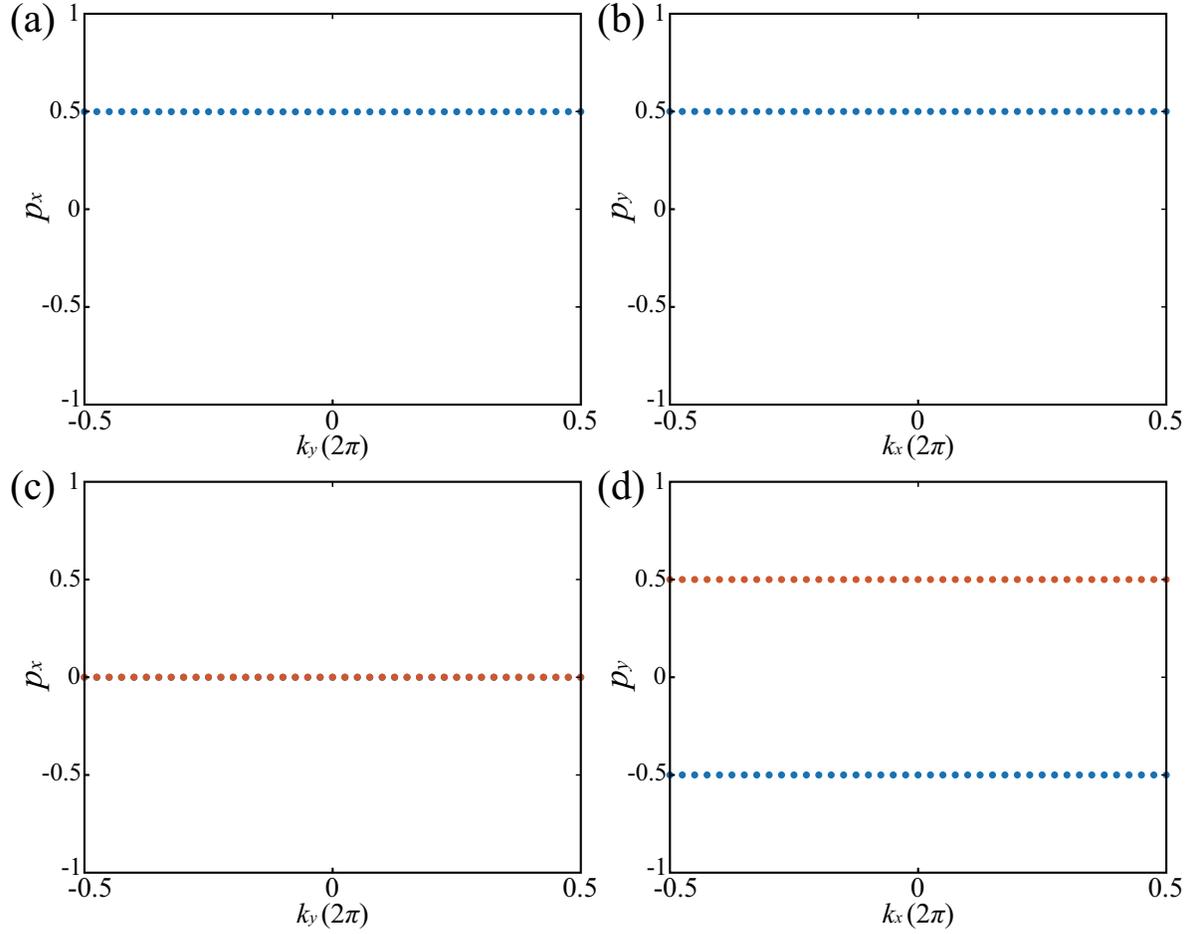}
    \caption{Polarizations of the 2D SSH model. (a)-(b) Wannier bands of one single occupied band considered in the $ k_{x} $-direction and $ k_{y} $-direction, respectively. (c)-(d) Wannier bands of two occupied bands considered in the $ k_{x} $-direction and $ k_{y} $-direction, respectively. Both the total polarizations are zero.}
    \label{fig7}
\end{figure}

In this section, we show the derivation of the acoustic model [Fig. \hyperref[fig5]{5(a)}]. According to the schematic dipicted in Fig. 1(a), we define the impedances between the intra- and inter-lattice cavities in the two directions as $ Z_{w}^{x}=i\omega L_{w}^{x} $, $ Z_{w}^{y}=i\omega L_{w}^{y} $, $ Z_{v}^{x}=i\omega L_{v}^{x} $ and $ Z_{v}^{y}=i\omega L_{v}^{y} $, respectively, and the intrinsic impedance of the cavities as $ Z_{c}=1/i\omega C $. Here, $ \omega = 2\pi f $ is the frequency of the wave, $ L_{p}^{q} $ $ ($ where $p = \{w, v\} $ and $ q = \{x, y\}) $ is the acoustic mass of the corresponding tube, and $ C $ is the 
acoustic capacitance of the cavity. We then introduce a Bloch wave function $ \lvert \textbf{u} \rangle = [u_{1}, u_{2}, u_{3}, u_{4}]^{\rm T} $ into the bulk lattice. Therefore, for the specific site,e.g., the 1-site, the flow of the wave component can be described as \cite{add1,add2}
\begin{equation}
    -\frac{u_{1}}{Z_{c}}=\frac{u_{1}-u_{3}^{x+}}{Z_{w}^{x}} + \frac{u_{1}-u_{4}^{y-}}{Z_{w}^{y}} + \frac{u_{1}-u_{3}^{x-}}{Z_{v}^{x}} + \frac{u_{1}-u_{4}^{y+}}{Z_{v}^{y}},
\label{eqB1}
\end{equation}
where the superscript $ ``x+" $ of the wave function indicates the nearest-neighbor site of 1-site in the positive $ x $-direction, and so on. For the other sites, we can obtain the similar forms. Further, we define $ w = -\omega^{2}Z_{c}/Z_{w} $ and $ v = -\omega^{2}Z_{c}/Z_{v} $. After Fourier expansion, the flows of $ \lvert \textbf{u}\rangle $ in the entire bulk lattice can be written in the reciprocal space as
\begin{equation}
    H(\textbf{k})\lvert \textbf{u}\rangle = \omega^{2}\lvert \textbf{u}\rangle,
    \label{eqB2}
\end{equation}
where
\begin{equation}
    H(\textbf{k}) = h(\textbf{k}) + \omega_{0}^{2}\tau_{0}\sigma_{0},
    \label{eqB3}
\end{equation}
is the rigorous Hamiltonian of the acoustic model, where $ \omega_{0}^{2} = w_{x}+w_{y}+v_{x}+v_{y} $ directly determines the frequency of the zero-energy states.

For the present acoustic resonance model, we can calculate all the parameters of the bulk lattice as 
\begin{equation}
\begin{aligned}
L_{w}&=\frac{\rho(l^{w}+1.7r)}{\pi r^{2}}, \\
L_{v}&=\rho(l^{v}+1.7r)/\pi r^{2}, \\
C &= a^{3}/\rho c^{2}. \\
\end{aligned}
\label{eqB4}
\end{equation}
Note that in the finite classical-wave systems, due to intrinsic chiral-symmetry breaking at the boundaries, there is always an on-site potential difference equaling to the inter-lattice hopping term $ v $ between the lattices at the boundaries and in the bulk \cite{add1, add2}, and therefore corrections should be added as stated in the main text. In addition, the soft boundary condition for preserving chiral symmetry at the boundaries of the structure, can be compensated by Helmholtz resonators with matched impedance in experimental realization \cite{Ni}. As a result, the bulk properties, i.e., hopping terms as well as the on-site potential in acoustic systems, can be directly obtained.


%

\newpage

\end{document}